\documentclass{llncs}
\bibliography{splncs03} 
\begin{document}
	\title{Perspectives on Surgical Data Science}
	\author{S. Swaroop Vedula\inst{1} and Masaru Ishii\inst{2} and Gregory D. Hager\inst{1}}
	\institute{Department of Computer Science, The Johns Hopkins University;	\email{vedula@jhu.edu, hager@cs.jhu.edu} \and Department of Otolaryngology-Head and Neck Surgery, The Johns Hopkins University School of Medicine; \email{mishii3@jhmi.edu}}
	\maketitle
	
	\begin{abstract}
The availability of large amounts of data together with advances in analytical techniques afford an opportunity to address difficult challenges in ensuring that healthcare is safe, effective, efficient, patient-centered, equitable, and timely. Surgical care and training stand to tremendously gain through surgical data science. Herein, we discuss a few perspectives on the scope and objectives for surgical data science.
	\end{abstract}
	
	\section{Introduction}
Surgical data science is an emerging discipline with the objective of enabling safe, effective, patient-centered, efficient, equitable, and timely surgical care by means of data acquisition, modeling, and analytics to empower clinical decision-making. Surgical data science relies for data capture technology upon mechanical, electrical, and electronics engineering, and for its analytic methods upon other data-intensive disciplines such as computer science, statistics, mathematics, information theory, and epidemiology. Data science involves extracting generalizable knowledge from large amounts of  data, which are often unstructured or complex, and yields actionable products for decision-making \cite{n0:dhar}. Surgical disciplines have yet to benefit from the data science revolution mainly because of limited data capture techniques. Surgical data science has now become feasible because of parallel developments in two domains --- technology to seamlessly capture surgical data at scale, and statistical techniques and computational algorithms to analyze large corpora of complex data from multiple sources.

While a data science approach is meaningful for any clinical discipline, we emphasize surgery because it is an indispensable and integral component of healthcare \cite{n1:ameara}. Globally, more than 234 million major surgical procedures are performed each year \cite{n2:weiser}. Still, about 5 billion humans lack access to high-quality surgical care \cite{n3:bmeara}. Delivering safe, effective, patient-centered, efficient, equitable, and timely surgical care is impeded by several challenges, some of which we propose can be addressed through a data science approach. In this article, we discuss a proposed scope for surgical data science and the challenges we anticipate in integrating data science into surgical care and education.
	
	\section{Scope for surgical data science}
	\paragraph{Surgical training} Effective and efficient surgical training is a necessity regardless of available resources. Although poor surgical skill is associated with a higher frequency of readmission, reoperation, and death \cite{n4:birk}, surgical education is inequitable, feedback provided to surgeons during training is inconsistent, and assessment of skill and competence remains unreliable \cite{n5:mock,n6:bell,n7:szasz}. Data-driven technologies for automated assessment of performance, diagnosis of skill deficits, targeted feedback through demonstration and examples, and individualized training are the next frontier for surgical education and credentialing across the globe. Technology for automated assessment may be focused on surgical technical skill and competence, non-technical skills, or clinical examination skills \cite{n7:szasz,n8:pugh}. But gaps exist in research on automated technologies for surgical training. First, most such research has emphasized technical skill although non-technical skills such as situation awareness, teamwork, and decision-making are critical for safe and effective surgical care \cite{n80:notss}. Second, technology to automate diagnosis of specific skill deficits and adaptive individualized feedback require learning from larger amounts of data than those typically utilized in research thus far. Third, algorithms to predict skill have yet to be translated into data products that can be integrated into surgical training curricula.

	\paragraph{Automated assistive technologies for surgical decision-making}
	Technology affords tremendous opportunity to enhance patient safety because it is seamlessly integrated into surgical care, for example, with laparoscopic, endoscopic, or robotic techniques. Technical errors accounted for patient injuries more than half the time in a study of surgical malpractice claims in the United States, of which a majority were manual errors and resulted in permanent disability or death \cite{n9:regen}. Automated assistive surgical technologies, such as an automated coach may mitigate certain technical errors. Manual coaching by an expert surgeon has been shown to improve surgical performance and reduce errors \cite{n10:bonrath}. Automated assistive surgical technologies may also serve other purposes such as surgical decision-making and context aware surgical systems that detect patient workflow and surgical activity. But automated intra-operative assistive surgical technologies are far from being adequately developed for deployment in the operating room because of two major limitations. First, although some algorithms perform with reasonable accuracy to recognize surgical phases, input from other stakeholders such as caregivers and patients is needed to transform the algorithms into tools with utility in the operating room. Second, techniques have yet to be developed that efficiently assimilate data from multiple sources (e.g., tool motion, video, environmental cues, physiologic monitoring) and enable development of applications using the extracted information. A collaborative data science approach has the potential to establish the role of assistive surgical technologies in pre-, intra-, and post-operative patient care.

	\paragraph{Variability and value of surgical patient care} 
	Variability in healthcare in general, and surgical care in particular, has had a large influence on evolution of policy. Landmark observations about geographic variation in patient outcomes has spurred an emphasis on quality of patient care \cite{n11:luft}. Most such efforts have focused on achieving certain benchamarks for a few patient outcomes at the population-level, e.g., 30-day mortality \cite{n11:luft}. But patient outcomes alone may not be adequate to develop a health system that delivers efficient, equitable, and timely care. Studying variability in patient care processes can yield improvements in efficiency and value of care, which is defined as outcomes achieved per monetary unit spent \cite{n12:value}. A data science approach can facilitate understanding and optimizing patient care processes in the pre- and post-operative settings because data in these contexts are often unstructured and available from disparate sources. Data-driven insights into optimizing patient care workflow before and after surgery can minimize variation in care patterns and patient outcomes. Algorithms to model the relationship between patients' clinical course following surgery and their outcomes can be used to develop tools for timely detection of patients at risk for poor outcomes, and efficiently allocating system resources.

	\paragraph{Effectiveness of surgical treatments for decision-making} 
	The effectiveness of surgical interventions is highly influenced by the skill with which they are delivered, unlike non-surgical treatments that involve simply ingesting a medication at a precisely titrated dose. Thus advances in surgical care through shared quantifiable knowledge require systematic, objective, and valid measures of surgical skill, both technical and non-technical. For example, setting up fair comparisons in randomized controlled trials or establishing registries to enable discovery of effective surgical treatments is impeded by lack of standardized, objective, and valid measures of skill and outcomes \cite{n13:cook}. Automated valid assessment of technical and non-technical surgical skill will also allow development of technological tools to support surgical decision-making in patient care, for example to individualize the choice of optimal surgical technique (laparoscopic vs. robotic) or treatment approach (e.g., use of mesh vs. natural tissue repair for a hernia).
	
	\section{Challenges to assimilate surgical data science}
A few major challenges that need to be addressed for surgical data science to reach fruition are mentioned here. First, surgical data are not routinely captured at scale. Integrating a culture of routine data collection into surgical care, and consistent data capture and processing infrastructure are necessary to promote surgical data science through uniform data repositories across institutions. Such repositories are essential to accelerate the pace of discoveries from data through collaborative analytics. Second, there is considerable heterogeneity in how surgical procedures are described across the world. Standardized ontologies to describe surgical procedures can not only facilitate collaborative analytics across data repositories but also promote discovery of shared knowledge across surgical procedures. In addition, uniform ontologies can catalyze the adoption of data products by healthcare decision-makers. Finally, current research using machine learning in surgery is geared towards optimizing algorithms to solve focused technical problems, e.g., surgical phase detection. Realistic data products, together with solutions for socio-technical issues of understanding how to integrate these new information sources into organizations in ways that allow their effective use, are needed to promote data science in healthcare decision-making.
	
	\section{Conclusion}
Surgical data science is a scientific discipline whose time has come. A data science approach to identify and address challenges in the surgical context will transform how data are used to advance patient care and surgical training. Educating key constituents, standardizing data infrastructure to maximize collective utility of disparate repositories, and promoting collaborations through dedicated funding mechanisms are necessary to realize the full potential of surgical data science.

\end{document}